\documentclass[aps,pra,preprintnumbers,showpacs,tightenlines]{revtex4}

\usepackage{amssymb}
\usepackage{amsmath}
\usepackage{graphicx}
\usepackage{epsfig}
\usepackage{subfigure}
\usepackage{amsfonts}
\usepackage{CJK}

\begin{document}

\title{A proposal for realizing a 3-qubit controlled-phase gate with
superconducting qubit systems coupled to a cavity}

\author{Chui-Ping Yang}
\email{yangcp@hznu.edu.cn}
\address{Department of Physics, Hangzhou Normal
University, Hangzhou, Zhejiang 310036, China}

\address{State Key Laboratory of Precision Spectroscopy, Department of Physics,
East China Normal University, Shanghai 200062, China}

\date{\today}

\begin{abstract}
We present a way to realize a 3-qubit quantum controlled-phase
gate with superconducting qubit systems coupled to a cavity. This
proposal does not require adjustment of the qubit level spacings
or identical qubit-cavity coupling constants. Moreover, since only a
resonant interaction is applied, the gate can be performed fast, within
$\sim $ 10 nanosecond. This proposal is quite general, which can be
applied to various types of superconducting qubits, atoms trapped in a
cavity, or quantum dots coupled to a resonator.
\end{abstract}

\pacs{03.67.Lx, 42.50.Dv, 85.25.Cp}\maketitle
\date{\today}

\begin{center}
\textbf{I. INTRODUCTION}
\end{center}

Superconducting qubit systems, including Cooper pair boxes,
Josephson junctions, and superconducting quantum interference
devices (SQUIDs), are among the most promising
candidates for scalable quantum computing [1]. In the past
decade, many theoretical methods for realizing a single-qubit gate
and a two-qubit gate with superconducting qubit systems have been
proposed [2-7]. Moreover, a two-qubit gate was experimentally
realized using superconduting qubit systems coupled through
capacitors [8-10], mutual inducance [11], or cavities [12,13].
However, the experimental realization of a 3-qubit
quantum gate with superconducting devices has not been reported so
far. Experimentally, a 3-qubit
controlled-phase gate has been realized in NMR quantum system and
a 3-qubit controlled NOT gate has been demonstrated with trapped
ions [14,15].

A 3-qubit controlled-phase (CP) gate is of significance, which has
applications in quantum information processing such as quantum network
circuit construction, error correction and quantum algorithms [16-19]. A
3-qubit CP gate can in principle be constructed using basic two-qubit
gates and single-qubit gates only (i.e., the conventional gate-decomposing
protocol). However, when using the conventional gate-decomposing protocol,
the gate operation is complicated because at least 25
steps of operations will be required, assuming that the realization of a
single-qubit gate or a two-qubit controlled phase gate requires a one-step
operation only [16] (see Fig. 1).

In this work, we focus on the physical realization of a 3-qubit CP
gate with superconducting qubit systems based on cavity QED
technique. Recently, cavity QED with superconducting qubits has
attracted considerable attention [20]. A cavity or resonator acts as a
quantum bus which can mediate long-distance, fast interaction between distant
superconducting qubit systems [4,12,21-26]. Based on cavity QED technique,
2-qubit quantum gates, 2-qubit quantum algorithms, 3-qubit quantum entanglement,
and quantum state transfer have been experimentally demonstrated with
superconducting qubits coupled to a cavity or resonator [12,13,27].

\begin{figure}[tbp]
\includegraphics[bb=68 335 548 674, width=12.6 cm, clip]{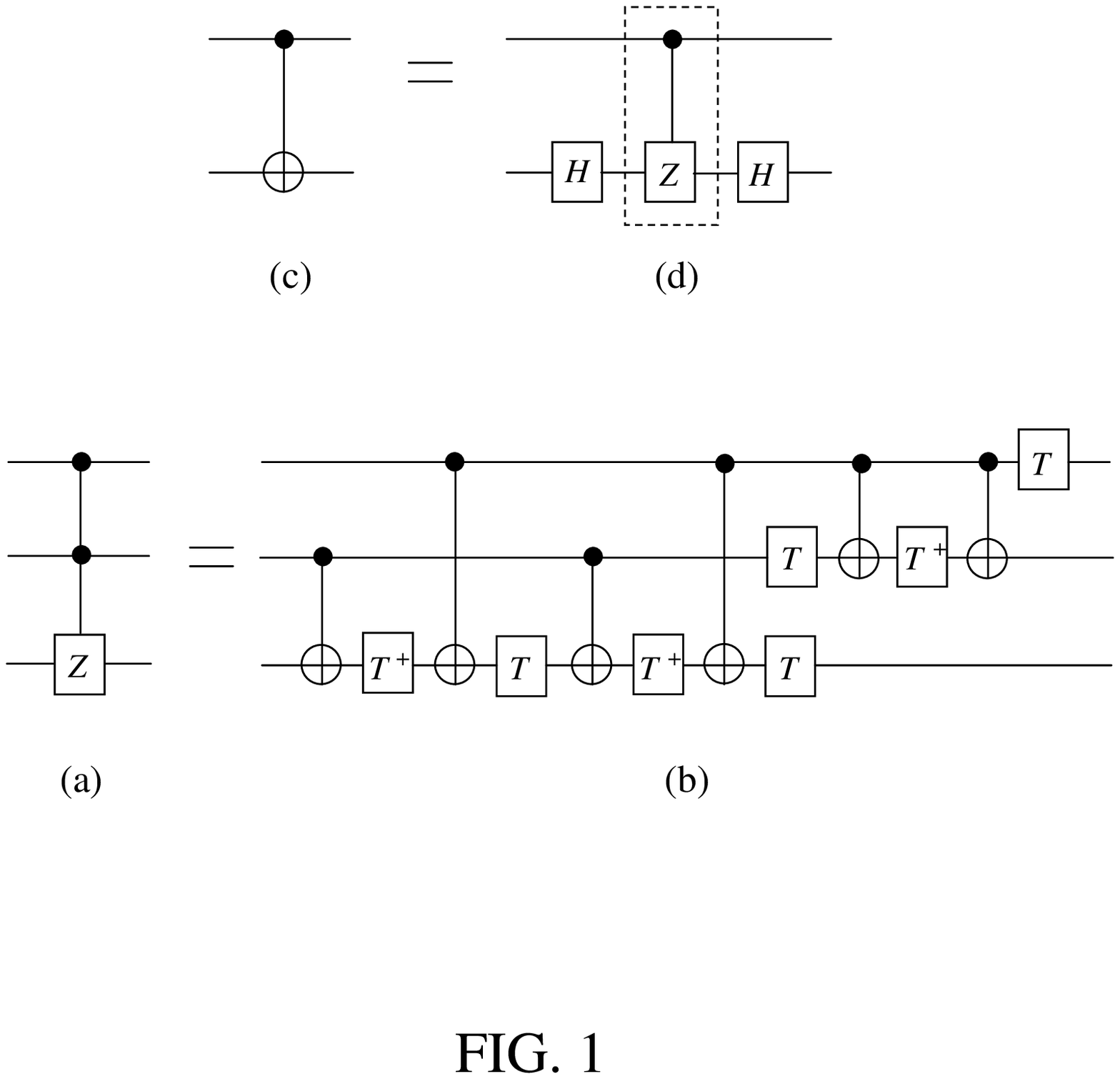} %
\vspace*{-0.08in}
\caption{(a) Schematic circuit of a 3-qubit CP gate with two control qubits
(linked to the filled circles) and a target qubit (at the bottom). Here, Z
represents a controlled-phase flip on the target qubit. The circuit in (a)
is equivalent to the circuit in (b), which consists of six 2-qubit
controlled NOT (CNOT) gates and seven single-qubit phase shift gates (i.e.,
the elements each containing a $T$ or $T^{+}$). (c) Schematic circuit for a
2-qubit CNOT gate with a control qubit (linked to the filled circle) and a
target qubit (at the bottom). The symbol $\oplus $ represents a controlled
NOT on the target qubit. The circuit in (c) is equivalent to the circuit in
(d), where the part enclosed in a dash-line box represents a 2-qubit CP gate
and each element containing $H$ stands for a single-qubit Hadamard gate. The
combination of circuits in (a), (b), (c), and (d) indicates that constructing a
3-qubit CP gate requires six 2-qubit CP gates, twelve single-qubit Hadamard
gates, and seven single-qubit phase shift gates. Namely, a total of 25 basic
gates are needed to build up a 3-qubit CP gate, by using the conventional
gate-decomposing protocol (for details, see reference [16]). Therefore,
at least 25 steps of operations will be required, assuming that the realization of a
single-qubit gate or a two-qubit controlled phase gate requires a one-step
operation only.}
\label{fig:1}
\end{figure}

In the following, we propose a way for realizing a 3-qubit CP
gate with superconducting qubit systems coupled to a cavity or
resonator (hereafter, we use the term cavity and resonator
interchangeably). As shown below, this proposal has the following
features: (a) there is no need for adjusting the qubit level
spacings during the gate operation, thus decoherence caused by
tuning the qubit level spacings is avoided; (b) no photon
detection is needed during the entire gate operation, thus the
effect of the photon-detection imperfection on the gate
performance is avoided; (c) identical qubit-cavity coupling
constants are not required, thus this proposal is tolerable to
the inevitable nonuniformity in device parameters, (d) since only
a resonant interaction is applied, the gate can be performed fast,
within $\sim $ 10 nanosecond.; and (e) this proposal only requires five
steps of operations. This proposal is quite general, which can be
applied to various types of superconducting qubits, atoms trapped in a
cavity, or quantum dots coupled to a resonator.

This paper is organized as follows. In Sec.~II, we briefly review
the basic theory of resonant qubit-cavity and qubit-pulse
interactions. In Sec.~III, we show how to realize a 3-qubit CP
gate with superconducting qubit systems coupled to a cavity or
resonator. In Sec.~IV, we briefly discuss possible experimental
implementation with superconducting qubit systems coupled to a
one-dimensional transmission line resonator. A concluding summary
is presented in Sec.~V.

\begin{figure}[tbp]
\includegraphics[bb=74 294 486 605, width=8.6 cm, clip]{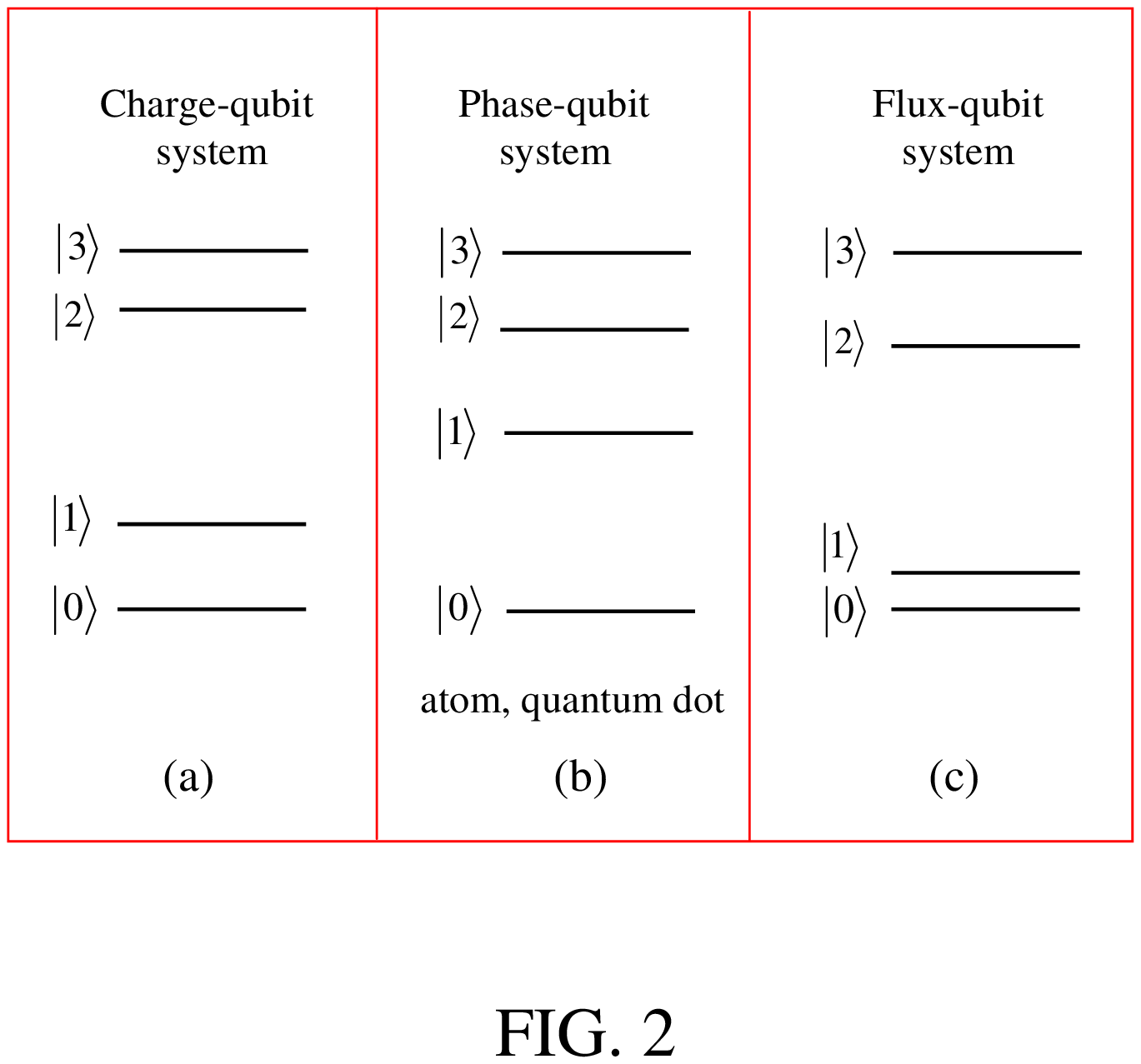} %
\vspace*{-0.08in} \caption{(Color online) Illustration of
four-level qubit systems. The energy eigenvalues for the four
levels $\left| 0\right\rangle$, $\left| 1\right\rangle$, $\left|
2\right\rangle$, and $\left| 3\right\rangle$ are denoted by $E_0$,
$E_1$, $E_2$, and $E_3$, respectively. In (a), the level spacings
satisfy $E_2-E_1>E_1-E_0,~E_3-E_2$; and $E_3-E_2<E_1-E_0$. In (b),
the level spacings satisfy $E_1-E_0>E_2-E_1>E_3-E_2$. In (c), the
level spacings meet $E_2-E_1>E_1-E_0,~E_3-E_2$; and
$E_3-E_2>E_1-E_0$. For the availability of the four levels in
superconducting systems, also see discussion in [33].}
\label{fig:2}
\end{figure}

\begin{center}
\textbf{II. BASIC THEORY}
\end{center}

The superconducting qubit systems considered in this work have
four levels shown in Fig.~2. Note that the four-level structure in
Fig.~2(a) applies to superconducting charge-qubit systems [28],
the one in Fig.~2(b) applies to phase-qubit systems [29,30], and
the one in Fig.~2(c) applies to flux-qubit systems [28,31]. In
addition, the four-level structure in Fig.~2(b) is also available
in natural atoms and quantum dots.

\textit{A. System-cavity resonant interaction.} Consider a system with four
levels as shown in Fig.~2. Suppose that the transition between the two
levels $\left| 2\right\rangle $ and $\left| 3\right\rangle$ is resonant with
the cavity mode while the transition between any other two levels is highly
detuned with (decoupled from) the cavity mode. In the interaction picture,
the interaction Hamiltonian of the system and the cavity mode is given by
(after the rotating-wave approximation)
\begin{equation}
H=\hbar g(a^{+}\sigma _{23}^{-}+\text{H.c.}),
\end{equation}
where $a^{+}$ and $a$ are the photon creation and annihilation operators of
the cavity mode, $g$ is the coupling constant between the cavity mode and
the $\left| 2\right\rangle \leftrightarrow \left| 3\right\rangle $
transition of the system, and $\sigma _{23}^{-}=\left| 2\right\rangle
\left\langle 3\right| $.

It is straightforward to show that the initial states $\left| 3\right\rangle
\left| 0\right\rangle _c$ and $\left| 2\right\rangle \left| 1\right\rangle
_c $ of the system and the cavity mode evolve as follows
\begin{eqnarray}
\left| 3\right\rangle \left| 0\right\rangle _c &\rightarrow &-i\sin \left(
gt\right) \left| 2\right\rangle \left| 1\right\rangle _c+\cos (gt)\left|
3\right\rangle \left| 0\right\rangle _c,  \nonumber \\
\left| 2\right\rangle \left| 1\right\rangle _c &\rightarrow &\cos \left(
gt\right) \left| 2\right\rangle \left| 1\right\rangle _c-i\sin \left(
gt\right) \left| 3\right\rangle \left| 0\right\rangle _c.
\end{eqnarray}
On the other hand, the state $\left| 0\right\rangle \left| 0\right\rangle _c$
remains unchanged under the Hamiltonian (1).

The coupling strength $g$ may vary with different systems due to different
level structures, non-uniform device parameters, and/or non-exact placement
of systems in the cavity. Therefore, in the gate operation below, we will
replace $g$ by $g_1$, $g_2$ and $g_3$ for systems $1$, $2$, and $3$,
respectively.

\textit{B. System-pulse resonant interaction.} Consider a system with four
levels as depicted in Fig.~2. Assume that the pulse is resonant with the
transition between the two levels $\left| i\right\rangle \leftrightarrow
\left| j\right\rangle $ of the system. Here, the level $\left|
i\right\rangle $ is a lower-energy level. The interaction Hamiltonian in the
interaction picture is given by
\begin{equation}
H_I=\hbar \left( \Omega _{ij}e^{i\phi }\left| i\right\rangle \left\langle
j\right| +\text{H.c.}\right) ,
\end{equation}
where $\Omega _{ij}$ and $\phi $ are the Rabi frequency and the initial
phase of the pulse, respectively. Based on the Hamiltonian (3), it is easy
to show that a pulse of duration $t$ results in the following rotation
\begin{eqnarray}
\left| i\right\rangle &\rightarrow &\cos \Omega _{ij}t\left| i\right\rangle
-ie^{-i\phi }\sin \Omega _{ij}t\left| j\right\rangle ,  \nonumber \\
\left| j\right\rangle &\rightarrow &\cos \Omega _{ij}t\left| j\right\rangle
-ie^{i\phi }\sin \Omega _{ij}t\left| i\right\rangle .
\end{eqnarray}
Note that the state transformation (4) can be completed within a very short
time, by increasing the pulse Rabi frequency $\Omega _{ij}$ (i.e., by
increasing the intensity of the pulse).

\begin{center}
\textbf{III. REALIZING A THREE-QUBIT CP GATE WITH SUPERCONDUCTING
QUBIT SYSTEMS COUPLED TO A CAVITY}
\end{center}

For three qubits, there are a total of eight computational
basis states, denoted by $\left| 000\right\rangle ,\left|
001\right\rangle ,...,\left| 111\right\rangle ,$ respectively. A
3-qubit CP gate results in the transformation $\left|
111\right\rangle \rightarrow -\left| 111\right\rangle $ but
nothing to the remaining seven computational basis states. Namely,
when the two control qubits (the first two qubits) are in the
state $\left| 1\right\rangle ,$ a
phase flip (i.e., $\left| 1\right\rangle \rightarrow -\left| 1\right\rangle $%
) happens to the state $\left| 1\right\rangle $ of the target
qubit (the last qubit). To realize this gate, let us consider
three superconducting qubit systems $1,$ $2,$ and $3.$ Each system
has four levels as depicted in Fig. 2. The two lowest levels
$\left| 0\right\rangle $ and $\left| 1\right\rangle $ of each
qubit system are used to represent the two logic states of a qubit
while the two higher-energy lowest levels $\left| 0\right\rangle $
and $\left| 1\right\rangle $ of each qubit system are employed for
the coherent manipulation of the quantum states. For simplicity,
we consider the four-level structure shown in Fig.~2(b) or Fig.~3
in our following discussion. For the purpose of the gate, we
denote the first (second) lowest level of system 1 or 2 as the
level $\left| 0\right\rangle $ ($\left|
1\right\rangle $) while the first (second) level of system 3 as the level $%
\left| 1\right\rangle $ ($\left| 0\right\rangle $) (see Fig~3). We
should mention that the gate operation procedure presented below
is applicable to the gate implementation using the four-level
configuration as depicted in Fig.~2(a) or Fig.~2(c).

The implementation of our gate below requires the resonant interaction
between the cavity mode and the $\left| 2\right\rangle \leftrightarrow
\left| 3\right\rangle $ transition of each qubit system. This condition can
be achieved by setting the level spacing between the two levels $\left|
2\right\rangle $ and $\left| 3\right\rangle $ to be the same for each qubit
system. Note that for superconducting qubit systems, by designing the qubit
systems appropriately, one can easily make the level spacing between certain
two levels (the two levels $\left| 2\right\rangle $ and $\left|
3\right\rangle $ here) to be identical [32], though it is diffcult to have
the level spacing between any two levels to be identical for each qubit
system due to nonuniformity of the system parameters. In addition, as shown
below, our gate realization requires that the cavity mode is highly detuned
(decoupled) from the transition between any other two levels of each system.
This condition can be achieved via adjustment of the qubit level spacings
before the gate operation. Note that for superconducting qubit systems, the
level spacings can be readily adjusted by changing the external parameters
(e.g., the external magnetic flux and gate voltage for superconducting
charge-qubit systems, the current bias or flux bias in the case of
superconducting phase-qubit systems and flux-qubit systems, see, e.g.
[28,29,33]).

\begin{figure}[tbp]
\includegraphics[bb=29 171 506 716, width=8.6 cm, clip]{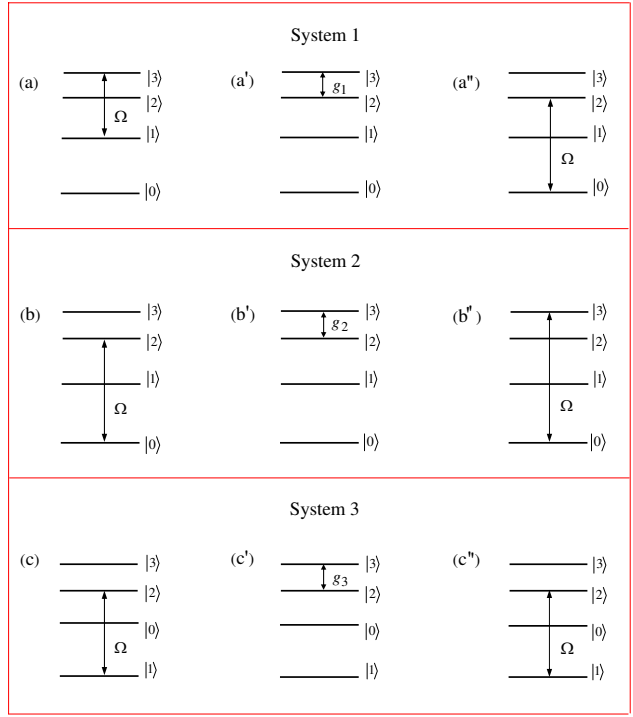} %
\vspace*{-0.08in}
\caption{(color online) Illustration of systems interacting with the cavity
mode or the pulses during the gate operation. For superconducting qubit
systems, the level spacing between certain two levels (the levels $\left|
2\right\rangle $ and $\left| 3\right\rangle $ here) can easily be made to be
the same for each qubit system (by appropriately designing the qubit
systems), such that the transition between these two levels for each qubit
is resonant with the cavity mode. For any other two levels, the
level-spacing difference for qubit systems 1, 2, and 3 is caused due to
nonuniformity of the qubit system parameters. The Rabi frequencies of the
pulses applied to the qubit systems 1, 2, and 3 can be set to be identical
by changing the intensity of the pulses.}
\label{fig:3}
\end{figure}

\begin{figure}[tbp]
\includegraphics[bb=119 249 383 620, width=8.6 cm, clip]{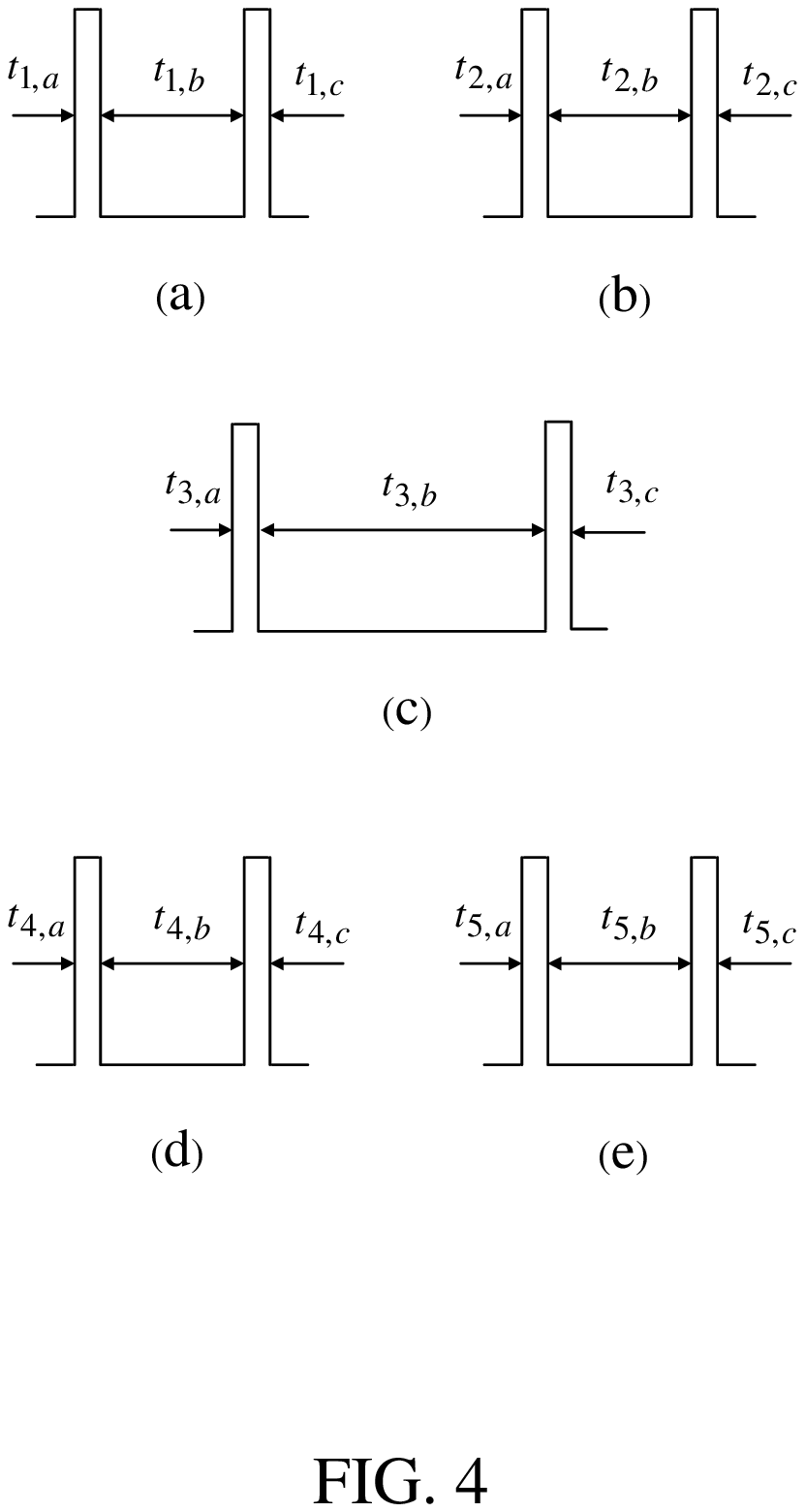} %
\vspace*{-0.08in}
\caption{Illustration of pulse sequences (from left to right) during the
gate operation. Figures (a) $\sim$ (e) correspond to steps (i) $\sim$ (v)
respectively, which are plotted for $g_1=g_2=g_3$. For each step of
operations, the pulse duration is much shorter than the waiting time $%
t_{i,b} $ ($i=1,2,,3,4,5$), due to $\Omega \gg g_1,g_2,g_3$.}
\label{fig:4}
\end{figure}

We now discuss how to implement our gate. The cavity mode is assumed to be
initially in the vacuum state $\left| 0\right\rangle _c.$ The procedure for
realizing the three-qubit CP gate is as follows:

Step (i): The operation of this step is: (a) Apply a pulse (with a frequency
$\omega =\omega _{31},$ a phase $\phi =-\frac \pi 2,$ and a duration $%
t_{1,a}=\frac \pi {2\Omega }$) to system $1$ [Fig.~3(a)], to transform the
state $\left| 1\right\rangle _1$ to $\left| 3\right\rangle _1$ as described
by Eq.~(4); (b) Wait for a time $t_{1,b}=\frac \pi {2g_1}$ to have the
cavity mode resonantly interacting with the $\left| 2\right\rangle
\leftrightarrow \left| 3\right\rangle $ transition of system $1$ [Fig.~3(a$%
^{\prime }$)], such that the state $\left| 3\right\rangle _1\left|
0\right\rangle _c$ is transformed to $-i\left| 2\right\rangle _1\left|
1\right\rangle _c$ as described by Eq. (2) while the state $\left|
0\right\rangle _1\left| 0\right\rangle _c$ remains unchanged; then (c) Apply
a pulse (with a frequency $\omega =\omega _{20},$ a phase $\phi =-\frac \pi
2,$ and a duration $t_{1,c}=\frac \pi {2\Omega }$) to system $1$ [Fig.~3(a$%
^{\prime \prime }$)], to transform the state $\left| 0\right\rangle _1$ to $%
\left| 2\right\rangle _1$ and the state $\left| 2\right\rangle _1$ to $%
-\left| 0\right\rangle _1$). The pulse sequence for this step of operation is shown in Fig.~4(a).

It can be seen that after the operation of this
step, the following transformation is obtained:

\begin{equation}
\begin{array}{c}
\left| 1\right\rangle _1\left| 0\right\rangle _c \\
\left| 0\right\rangle _1\left| 0\right\rangle _c
\end{array}
\stackrel{\left( a\right) }{\rightarrow }
\begin{array}{c}
\left| 3\right\rangle _1\left| 0\right\rangle _c \\
\left| 0\right\rangle _1\left| 0\right\rangle _c
\end{array}
\stackrel{\left( b\right) }{\rightarrow }
\begin{array}{c}
-i\left| 2\right\rangle _1\left| 1\right\rangle _c \\
\left| 0\right\rangle _1\left| 0\right\rangle _c
\end{array}
\stackrel{\left( c\right) }{\rightarrow }
\begin{array}{c}
i\left| 0\right\rangle _1\left| 1\right\rangle _c \\
\left| 2\right\rangle _1\left| 0\right\rangle _c
\end{array}
.
\end{equation}

Step (ii): The operation for this step is: (a) Apply a pulse (with a
frequency $\omega =\omega _{20},$ a phase $\phi =-\frac \pi 2,$ and a
duration $t_{2,a}=\frac \pi {2\Omega }$) to system $2$ [Fig.~3(b)], to
transform the state $\left| 0\right\rangle _2$ to $\left| 2\right\rangle _2$%
; (b) Wait for a time $t_{2,b}=\frac \pi {2g_2}$ to have the cavity mode
resonantly interacting with the $\left| 2\right\rangle \leftrightarrow
\left| 3\right\rangle $ transition of system $2$ [Fig.~3(b$^{\prime }$)],
such that the state $\left| 2\right\rangle _2\left| 1\right\rangle _c$ is
transformed to $-i\left| 3\right\rangle _2\left| 0\right\rangle _c$ while
the states $\left| 1\right\rangle _2\left| 0\right\rangle _c,$ $\left|
2\right\rangle _2\left| 0\right\rangle _c$ and $\left| 1\right\rangle
_2\left| 1\right\rangle _c$ remain unchanged; then (c) Apply a pulse (with a
frequency $\omega =\omega _{30},$ a phase $\phi =\pi ,$ and a duration $%
t_{2,c}=\frac \pi {2\Omega }$) to system $2$ [Fig.~3(b$^{\prime \prime }$)],
to transform the state $\left| 3\right\rangle _2$ to $i\left| 0\right\rangle
_2.$ The pulse sequence for this step of operation is shown in Fig.~4(b).

One can see that after the operation of this step, the following
transformation is achieved:

\begin{equation}
\begin{array}{c}
\left| 0\right\rangle _2\left| 1\right\rangle _c \\
\left| 1\right\rangle _2\left| 1\right\rangle _c \\
\left| 0\right\rangle _2\left| 0\right\rangle _c \\
\left| 1\right\rangle _2\left| 0\right\rangle _c
\end{array}
\stackrel{\left( a\right) }{\rightarrow }
\begin{array}{c}
\left| 2\right\rangle _2\left| 1\right\rangle _c \\
\left| 1\right\rangle _2\left| 1\right\rangle _c \\
\left| 2\right\rangle _2\left| 0\right\rangle _c \\
\left| 1\right\rangle _2\left| 0\right\rangle _c
\end{array}
\stackrel{\left( b\right) }{\rightarrow }
\begin{array}{c}
-i\left| 3\right\rangle _2\left| 0\right\rangle _c \\
\left| 1\right\rangle _2\left| 1\right\rangle _c \\
\left| 2\right\rangle _2\left| 0\right\rangle _c \\
\left| 1\right\rangle _2\left| 0\right\rangle _c
\end{array}
\stackrel{\left( c\right) }{\rightarrow }
\begin{array}{c}
\left| 0\right\rangle _2\left| 0\right\rangle _c \\
\left| 1\right\rangle _2\left| 1\right\rangle _c \\
\left| 2\right\rangle _2\left| 0\right\rangle _c \\
\left| 1\right\rangle _2\left| 0\right\rangle _c
\end{array}
.
\end{equation}

Step (iii): The operation for this step is: (a) Apply a pulse (with a
frequency $\omega =\omega _{21},$ a phase $\phi =-\frac \pi 2,$ and a
duration $t_{3,a}=\frac \pi {2\Omega }$) to system $3$ [Fig.~3(c)], to
transform the state $\left| 1\right\rangle _3$ to $\left| 2\right\rangle _3$%
; (b) Wait for a time $t_{3,b}=\frac \pi {g_3}$ to have the cavity mode
resonantly interacting with the $\left| 2\right\rangle \leftrightarrow
\left| 3\right\rangle $ transition of system $3$ [Fig.~3(c$^{\prime }$)],
such that the state $\left| 2\right\rangle _3\left| 1\right\rangle _c$
changes to $-\left| 2\right\rangle _3\left| 1\right\rangle _c$ while the
states $\left| 2\right\rangle _3\left| 0\right\rangle _c,$ $\left|
0\right\rangle _3\left| 0\right\rangle _c,$ and $\left| 0\right\rangle
_3\left| 1\right\rangle _c$ remain unchanged; then (c) Apply a pulse (with a
frequency $\omega =\omega _{21},$ a phase $\phi =\frac \pi 2,$ and a
duration $t_{3,c}=\frac \pi {2\Omega }$) to system $3$ [Fig.~3(c$^{\prime
\prime }$)], to transform the state $\left| 2\right\rangle _3$ back to $%
\left| 1\right\rangle _3.$ The pulse sequence for this step of operation is shown in Fig.~4(c).

After the operation of this step, we obtain the following transformation:

\begin{equation}
\begin{array}{c}
\left| 0\right\rangle _3\left| 0\right\rangle _c \\
\left| 1\right\rangle _3\left| 0\right\rangle _c \\
\left| 0\right\rangle _3\left| 1\right\rangle _c \\
\left| 1\right\rangle _3\left| 1\right\rangle _c
\end{array}
\stackrel{\left( a\right) }{\rightarrow }
\begin{array}{c}
\left| 0\right\rangle _3\left| 0\right\rangle _c \\
\left| 2\right\rangle _3\left| 0\right\rangle _c \\
\left| 0\right\rangle _3\left| 1\right\rangle _c \\
\left| 2\right\rangle _3\left| 1\right\rangle _c
\end{array}
\stackrel{\left( b\right) }{\rightarrow }
\begin{array}{c}
\left| 0\right\rangle _3\left| 0\right\rangle _c \\
\left| 2\right\rangle _3\left| 0\right\rangle _c \\
\left| 0\right\rangle _3\left| 1\right\rangle _c \\
-\left| 2\right\rangle _3\left| 1\right\rangle _c
\end{array}
\stackrel{\left( c\right) }{\rightarrow }
\begin{array}{c}
\left| 0\right\rangle _3\left| 0\right\rangle _c \\
\left| 1\right\rangle _3\left| 0\right\rangle _c \\
\left| 0\right\rangle _3\left| 1\right\rangle _c \\
-\left| 1\right\rangle _3\left| 1\right\rangle _c
\end{array}
.
\end{equation}

The purpose of the last two steps presented below is to obtain the reverse
transformations described by Eqs. (5) and (6). The operations for these two
steps are as follows:

Step (iv): (a) Apply a pulse (with a frequency $\omega =\omega _{30},$ a
phase $\phi =\pi ,$ and a duration $t_{4,a}=\frac \pi {2\Omega }$) to system
$2$ [Fig.~3(b$^{\prime \prime }$)], to transform the state $\left|
0\right\rangle _2$ to $i\left| 3\right\rangle _2$; (b) Wait for a time $%
t_{4,b}=\frac \pi {2g_2}$ to have the cavity mode resonant interact with the
$\left| 2\right\rangle \leftrightarrow \left| 3\right\rangle $ transition of
system $2$ [Fig.~3(b$^{\prime }$)], such that the state $\left|
3\right\rangle _2\left| 0\right\rangle _c$ is transformed to $-i\left|
2\right\rangle _2\left| 1\right\rangle _c$ while the states $\left|
1\right\rangle _2\left| 0\right\rangle _c,$ $\left| 2\right\rangle _2\left|
0\right\rangle _c$ and $\left| 1\right\rangle _2\left| 1\right\rangle _c$
remain unchanged; then (c) Apply a pulse (with a frequency $\omega =\omega
_{20},$ a phase $\phi =\frac \pi 2,$ and a duration $t_{4,c}=\frac \pi
{2\Omega }$) to system $2$ [Fig.~3(b)], to transform the state $\left|
2\right\rangle _2$ to $\left| 0\right\rangle _2.$ The pulse sequence for this step
of operation is shown in Fig.~4(d).

One can see that after the operation of this step, the following
transformation is achieved:

\begin{equation}
\begin{array}{c}
\left| 0\right\rangle _2\left| 0\right\rangle _c \\
\left| 1\right\rangle _2\left| 1\right\rangle _c \\
\left| 2\right\rangle _2\left| 0\right\rangle _c \\
\left| 1\right\rangle _2\left| 0\right\rangle _c
\end{array}
\stackrel{\left( a\right) }{\rightarrow }
\begin{array}{c}
i\left| 3\right\rangle _2\left| 0\right\rangle _c \\
\left| 1\right\rangle _2\left| 1\right\rangle _c \\
\left| 2\right\rangle _2\left| 0\right\rangle _c \\
\left| 1\right\rangle _2\left| 0\right\rangle _c
\end{array}
\stackrel{\left( b\right) }{\rightarrow }
\begin{array}{c}
\left| 2\right\rangle _2\left| 1\right\rangle _c \\
\left| 1\right\rangle _2\left| 1\right\rangle _c \\
\left| 2\right\rangle _2\left| 0\right\rangle _c \\
\left| 1\right\rangle _2\left| 0\right\rangle _c
\end{array}
\stackrel{\left( c\right) }{\rightarrow }
\begin{array}{c}
\left| 0\right\rangle _2\left| 0\right\rangle _c \\
\left| 1\right\rangle _2\left| 1\right\rangle _c \\
\left| 0\right\rangle _2\left| 0\right\rangle _c \\
\left| 1\right\rangle _2\left| 0\right\rangle _c
\end{array}
.
\end{equation}

Step (v): (a) Apply a pulse (with a frequency $\omega =\omega _{20},$ a
phase $\phi =\frac \pi 2,$ and a duration $t_{5,a}=\frac \pi {2\Omega }$) to
system $1$ [Fig.~3(a$^{\prime \prime }$)], to transform $\left|
0\right\rangle _1$ to $-\left| 2\right\rangle _1$ and $\left| 2\right\rangle
_1$ to $\left| 0\right\rangle _1$; (b) Wait for a time $t_{5,b}=\frac \pi
{2g_1}$ to have the cavity mode resonantly interacting with the $\left|
2\right\rangle \leftrightarrow \left| 3\right\rangle $ transition of system $%
1$ [Fig.~3(a$^{\prime }$)], such that the state $\left| 2\right\rangle
_1\left| 1\right\rangle _c$ is transformed to $-i\left| 3\right\rangle
_1\left| 0\right\rangle _c$ as described by Eq. (2) while the state $\left|
0\right\rangle _1\left| 0\right\rangle _c$ remains unchanged; then (c) Apply
a pulse (with a frequency $\omega =\omega _{31},$ a phase $\phi =-\frac \pi
2,$ and a duration $t_{5,c}=\frac \pi {2\Omega }$) to system $1$
[Fig.~3(a)], to transform the state $\left| 3\right\rangle _1$ to $-\left|
1\right\rangle _1.$ The pulse sequence for this step of operation is shown in Fig.~4(e).

It can be seen that after the operation of this step, the following
transformation is obtained:

\begin{equation}
\begin{array}{c}
\left| 0\right\rangle _1\left| 1\right\rangle _c \\
\left| 2\right\rangle _1\left| 0\right\rangle _c
\end{array}
\stackrel{\left( a\right) }{\rightarrow }
\begin{array}{c}
-\left| 2\right\rangle _1\left| 1\right\rangle _c \\
\left| 0\right\rangle _1\left| 0\right\rangle _c
\end{array}
\stackrel{\left( b\right) }{\rightarrow }
\begin{array}{c}
i\left| 3\right\rangle _1\left| 0\right\rangle _c \\
\left| 0\right\rangle _1\left| 0\right\rangle _c
\end{array}
\stackrel{\left( c\right) }{\rightarrow }
\begin{array}{c}
-i\left| 1\right\rangle _1\left| 0\right\rangle _c \\
\left| 0\right\rangle _1\left| 0\right\rangle _c
\end{array}
.
\end{equation}

Based on the results (5-9) obtained above, we can find the following state
evolution of the whole system after each step of the above operations:
\[
\begin{array}{c}
\left| 100\right\rangle \left| 0\right\rangle _c \\
\left| 101\right\rangle \left| 0\right\rangle _c \\
\left| 110\right\rangle \left| 0\right\rangle _c \\
\left| 111\right\rangle \left| 0\right\rangle _c
\end{array}
\stackrel{\text{Step(i)}}{\longrightarrow }
\begin{array}{c}
i\left| 000\right\rangle \left| 1\right\rangle _c \\
i\left| 001\right\rangle \left| 1\right\rangle _c \\
i\left| 010\right\rangle \left| 1\right\rangle _c \\
i\left| 011\right\rangle \left| 1\right\rangle _c
\end{array}
\stackrel{\text{Step(ii)}}{\longrightarrow }
\begin{array}{c}
i\left| 000\right\rangle \left| 0\right\rangle _c \\
i\left| 001\right\rangle \left| 0\right\rangle _c \\
i\left| 010\right\rangle \left| 1\right\rangle _c \\
i\left| 011\right\rangle \left| 1\right\rangle _c
\end{array}
\stackrel{\text{Step(iii)}}{\longrightarrow }
\begin{array}{c}
i\left| 000\right\rangle \left| 0\right\rangle _c \\
i\left| 001\right\rangle \left| 0\right\rangle _c \\
i\left| 010\right\rangle \left| 1\right\rangle _c \\
-i\left| 011\right\rangle \left| 1\right\rangle _c
\end{array}
\]
\begin{equation}
\stackrel{\text{Step(iv)}}{\longrightarrow }
\begin{array}{c}
i\left| 000\right\rangle \left| 1\right\rangle _c \\
i\left| 001\right\rangle \left| 1\right\rangle _c \\
i\left| 010\right\rangle \left| 1\right\rangle _c \\
-i\left| 011\right\rangle \left| 1\right\rangle _c
\end{array}
\stackrel{\text{Step(v)}}{\longrightarrow }
\begin{array}{c}
\left| 100\right\rangle \left| 0\right\rangle _c \\
\left| 101\right\rangle \left| 0\right\rangle _c \\
\left| 110\right\rangle \left| 0\right\rangle _c \\
-\left| 111\right\rangle \left| 0\right\rangle _c
\end{array}
.
\end{equation}
Here and below, $\left| ijk\right\rangle $ is abbreviation of the state $%
\left| i\right\rangle _1\left| j\right\rangle _2\left| k\right\rangle _3$ of
systems ($1,2,3$) with $i,j,k\in \{0,1,2,3,4\}$. This result (10)
demonstrates that a phase flip happens to the state $\left| 111\right\rangle
$ after the operations above.

Eq. (10) shows that during each step of operations, the two states $\left|
0\right\rangle $ and $\left| 1\right\rangle $ of each of two irrelevant
qubit systems remain unchanged. For instance, one can see from Eq. (10) that
during the operation of step (i) on qubit 1, the two states $\left|
0\right\rangle $ and $\left| 1\right\rangle $ of qubit system 2 or qubit
system 3 remain unchanged. The reason for this is that the cavity mode was
assumed to be resonant with the $\left| 2\right\rangle \leftrightarrow
\left| 3\right\rangle $ transition of each qubit but highly detuned
(decoupled) from the transition between any other two levels. Thus, when a
qubit is in the state $\left| 0\right\rangle $ or $\left| 1\right\rangle $,
the qubit is decoupled from the cavity mode and thus the states $\left|
0\right\rangle $ and $\left| 1\right\rangle $ of this qubit are not affected
by the cavity mode though a photon is populated in the cavity.

On the other hand, based on the results (5-9) given above, it is easy to see
that for the other four initial states $\left| 000\right\rangle ,\left|
001\right\rangle ,\left| 010\right\rangle ,$ and $\left| 011\right\rangle $
of the qubit systems, the states of the whole system after each step of the
above operations are listed below:

\[
\begin{array}{c}
\left| 000\right\rangle \left| 0\right\rangle _c \\
\left| 001\right\rangle \left| 0\right\rangle _c \\
\left| 010\right\rangle \left| 0\right\rangle _c \\
\left| 011\right\rangle \left| 0\right\rangle _c
\end{array}
\stackrel{\text{Step(i)}}{\longrightarrow }
\begin{array}{c}
\left| 200\right\rangle \left| 0\right\rangle _c \\
\left| 201\right\rangle \left| 0\right\rangle _c \\
\left| 210\right\rangle \left| 0\right\rangle _c \\
\left| 211\right\rangle \left| 0\right\rangle _c
\end{array}
\stackrel{\text{Step(ii)}}{\longrightarrow }
\begin{array}{c}
\left| 220\right\rangle \left| 0\right\rangle _c \\
\left| 221\right\rangle \left| 0\right\rangle _c \\
\left| 210\right\rangle \left| 0\right\rangle _c \\
\left| 211\right\rangle \left| 0\right\rangle _c
\end{array}
\stackrel{\text{Step(iii)}}{\longrightarrow }
\begin{array}{c}
\left| 220\right\rangle \left| 0\right\rangle _c \\
\left| 221\right\rangle \left| 0\right\rangle _c \\
\left| 210\right\rangle \left| 0\right\rangle _c \\
\left| 211\right\rangle \left| 0\right\rangle _c
\end{array}
\]

\begin{equation}
\stackrel{\text{Step(iv)}}{\longrightarrow }
\begin{array}{c}
\left| 200\right\rangle \left| 0\right\rangle _c \\
\left| 201\right\rangle \left| 0\right\rangle _c \\
\left| 210\right\rangle \left| 0\right\rangle _c \\
\left| 211\right\rangle \left| 0\right\rangle _c
\end{array}
\stackrel{\text{Step(v)}}{\longrightarrow }
\begin{array}{c}
\left| 000\right\rangle \left| 0\right\rangle _c \\
\left| 001\right\rangle \left| 0\right\rangle _c \\
\left| 010\right\rangle \left| 0\right\rangle _c \\
\left| 011\right\rangle \left| 0\right\rangle _c
\end{array}
,
\end{equation}
which shows that the cavity mode remains in the vacuum state $\left|
0\right\rangle _c$ during the entire operation, the four states $\left|
000\right\rangle ,\left| 001\right\rangle ,\left| 010\right\rangle $ and $%
\left| 011\right\rangle $ of the qubit systems undergo time evolution
(induced by the applied pulses only) but return to their original states
after the last step of operation. Note that the state $\left| 2\right\rangle
$ of qubit system 1, 2, or 3 does not change when no photon is populated in
the cavity, due to the energy conservation. From Eqs. (10) and (11), it can
be concluded that after the above process, a 3-qubit CP gate was implemented
with three systems (i.e., the controlled systems $1$ and $2,$ as well as the
target system $3$) while the cavity mode returns to its original vacuum
state.

Several points need to be addressed as follows:

(i) From the description of operations above, it can be seen that due to the
use of the four levels, the states $\left| 0\right\rangle $ and $\left|
1\right\rangle $ of qubit systems not involved in the operation are not
affected by the cavity mode, no matter whether or not a photon is populated
in the cavity. We note that when systems with two or three energy levels are
used, the decoupling of other systems from the cavity mode can not be made
during the operation performed on any one of the systems.

(ii) For certain kinds of superconducting qubit systems (e.g., flux qubit
systems), the decay of the level $\left| 1\right\rangle $ of each system is
avoided when the transition between the two lowest levels is forbidden due
to the optical selection rules [31], or it can be suppressed by increasing
the potential barrier between the two lowest levels $\left| 0\right\rangle $
and $\left| 1\right\rangle $ [4,29,30,34]

(iii) For simplicity, we considered the identical Rabi frequency $\Omega$
for each pulse during the gate operation above. Note that this requirement
is unnecessary. The Rabi frequency for each pulse can be different and thus
the pulse durations for each step of operations above can be adjusted
accordingly.

(iv) To have the effect of the system-cavity resonant interaction on the state
transformation induced by the pulse negligible, the pulse Rabi
frequency $\Omega $ needs to be set such that $\Omega \gg g_1,g_2,g_3$.

\begin{figure}[tbp]
\includegraphics[bb=40 221 563 699, width=8.2 cm, clip]{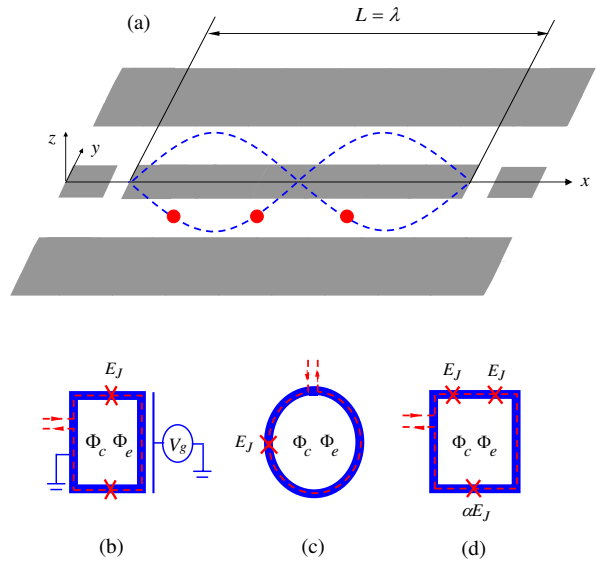} %
\vspace*{-0.08in}
\caption{(color online) (a) Setup for three superconducting qubit systems
(red dots) and a (grey) standing-wave one-dimensional coplanar waveguide
resonator. $\lambda$ is the wavelength of the resonator mode, and $L$ is the
length of the resonator. The two (blue) curved lines represent the standing
wave magnetic field in the $z$-direction. Each qubit system (a red dot)
could be a superconducting charge-qubit system as depicted in (b),
flux-biased phase-qubit system in (c), and flux-qubit system in (d). $E_{J}$
is the Josephson junction energy ($0.6<\alpha<0.8$) and $V_g$ is the gate
voltage. The qubit systems are placed at locations where the magnetic fields
are the same to achieve an identical coupling strength for each qubit
system. The superconducting loop of each qubit system, which is a large
square for (b) and (d) while a large circle for (c), is located in the plane
of the resonator between the two lateral ground planes (i.e., the $x$-$y$
plane). For each qubit system, the external magnetic flux $\Phi_c$ through
the superconducting loop for each qubit system is created by the magnetic
field threading the superconducting loop. A classical magnetic pulse is
applied to each qubit system through an \textit{ac} flux $\Phi_e$ threading
the qubit superconducting loop, which is created by an \textit{ac} current
loop (i.e., the red dashed-line loop) placed on the qubit loop. The pulse
frequency and intensity can be adjusted by changing the frequency and
intensity of the \textit{ac} loop current.}
\label{fig:5}
\end{figure}

\begin{center}
\textbf{IV. POSSIBLE EXPERIMENTAL IMPLEMENTATION}
\end{center}

As shown above, it can be found that the total operation time is
given by
\begin{eqnarray}
\tau &=&\sum_{i=1}^5\left( t_{i,a}+t_{i,b}+t_{i,c}\right)  \nonumber \\
\ &=&\pi /g_1+\pi /g_2+\pi /g_3+5\pi /\Omega .
\end{eqnarray}
The $\tau $ should be much shorter than the energy relaxation time
$\gamma _{3r}^{-1}$ and dephasing time $\gamma _{3p}^{-1}$ of the
level $\left| 3\right\rangle $ (note that the levels $\left|
1\right\rangle $ and $\left| 2\right\rangle $ have a longer
decoherence time than the level $\left| 3\right\rangle $), such
that decoherence, caused due to spontaneous decay and dephasing
process of the qubit systems, is negligible during the operation.
And, the $\tau $ needs to be much shorter than the lifetime of the
cavity photon, which is given by $\kappa ^{-1}=Q/2\pi \nu _c,$
such that
the decay of the cavity photon can be neglected during the operation. Here, $%
Q$ is the (loaded) quality factor of the cavity and $\nu _c$ is
the cavity field frequency. To obtain these requirements, one can
design the qubit systems to have sufficiently long energy
relaxation time and dephasing time, such that $\tau \ll $ $\gamma
_{3r}^{-1},\gamma _{3p}^{-1};$ and choose a high-$Q$ cavity such
that $\tau \ll \kappa ^{-1}.$

For the sake of definitiveness, let us consider the experimental
possibility of realizing the 3-qubit CP gate, using three
identical superconducting qubit systems coupled to a resonator
[Fig.~5(a)]. Each qubit system could be a superconducting
charge-qubit system [Fig.~5(b)], flux-qubit system [Fig.~5(c)], or
flux-biased phase-qubit system [Fig.~5(d)]. As a rough estimate,
assume $g_1\approx g_2\approx g_3=g$, and $g/{2\pi }\sim 220$ MHz,
which could be reached for a superconducting qubit system coupled
to a one-dimensional standing-wave CPW (coplanar waveguide)
transmission resonator [27]. With the choice of $\Omega \sim 10g,$ one has $\tau \sim 8$ $%
n$s, much shorter than $\min \{\gamma _{3r}^{-1},\gamma _{3p}^{-1}\}\sim 1$ $%
\mu $s [29,35]. In addition, consider a resonator with frequency $\nu _c\sim
5$ GHz (e.g., Ref.~[13]) and $Q\sim 5\times 10^4$, we have $\kappa ^{-1}\sim
1.6$ $\mu $s, which is much longer than the operation time $\tau $ here.
Note that superconducting coplanar waveguide resonators with a (loaded)
quality factor $Q\sim 10^6$ have been experimentally demonstrated [36,37].
We remark that further investigation is needed for each particular
experimental setup. However, this requires a rather lengthy and complex
analysis, which is beyond the scope of this theoretical work.

\begin{center}
\textbf{V. CONCLUSION}
\end{center}

Before conclusion, we should mention two previous proposals on
multiqubit-controlled phase gates [38,39], which are relevant to
this work. However, we note that the present proposal is quite
different from the ones in [38,39] as follows:

(i) The previous proposal in [38] requires using {\it five-level}
qubits while our present proposal employs {\it four-level} qubits
only. Since one more level is used, the former is more challenging
in experiments when compared with the latter. Furthermore, the
proposal in [38] is based on the adiabatic passage technique while
ours is based on the resonant interaction only. As is well known,
the adiabatic passage requires slowly changing the Rabi
frequencies of the pulses applied. Hence, the gate speed for the
proposal in [38] is far slower than that using our present
proposal.

(ii) The previous proposal in [39] requires adjusting the qubit
level spacings {\it during the gate operation} while as shown
above our present proposal does not need adjustment of the qubit
level spacings during the gate operation. Thus, decoherence caused
due to adjustment of the qubit level spacings is avoided by our
present proposal. Moreover, adjusting the qubit level spacings
during the gate operation is undesirable in experiments, which
however does not apply to our present proposal. Hence, the present
proposal is much improved when compared with the proposal in [39].

In summary, we have presented a way to realize a 3-qubit
controlled-phase gate with four-level superconducting qubit
systems in cavity QED. As shown above, this proposal has the
following advantages: (i) No adjustment of the level spacings of
qubit systems during the entire operation is needed, thus
decoherence caused due to the adjustment of the level spacings is
avoided in this proposal; (ii) The coupling constants of each
system with the cavity are not required to be identical, which
makes neither identical qubits nor exact placement of qubits to be
required by this proposal; (iii) No photon detection is needed
during the entire gate operation, and thus the effect of the
photon-detection imperfection on the gate performance is avoided;
(iv) Because only resonant interactions are used, the gate can be
performed fast within $\sim $ 10 nanosecond; and (v) The gate
realization requires five steps of operations only. Finally, it is
noted that this proposal is quite general, which can be applied to
other physical systems, such as atoms trapped in a cavity or
quantum dots coupled to a resonator.

\begin{center}
\textbf{ACKNOWLEDGMENTS}
\end{center}

This work was supported in part by the National Natural Science Foundation of China
under Grant No. 11074062, the Zhejiang Natural Science Foundation under
Grant No. Y6100098, the Open Fund from the SKLPS of ECNU, and the funds from Hangzhou Normal University.

\end{document}